# Stabilizing low symmetry-based functions of materials at room temperature through isosymmetric electronic bistability

Francisco Javier Valverde-Muñoz,[a] Ricardo Guillermo Torres Ramírez,[a] Elzbieta Trzop,[a,b] Thierry Bataille,[c] Nathalie Daro,[d] Dominique Denux,[d] Philippe Guionneau,[d] Hervé Cailleau,[a] Guillaume Chastanet,[d] Boris Le Guennic,[c] Eric Collet*[a,b,e]

Symmetry-breaking is pivotal for controlling ferroelectric, ferroelastic and/or ferromagnetic functions of materials, which enables applications in sensors, memories, transducers or actuators. Commonly, ferroic phases emerge from descending symmetry-breaking, as the laws of thermodynamics dictate that the ordered low entropy phases form at low temperature, which limits practical applications of many materials at room temperature. Rare examples of ascending symmetry-breakings have been observed, but the driving force remains often unclear. Here, we report on a ferroelastic symmetry-breaking occurring at high temperature in a spin-crossover material, studied by magnetic, DSC and X-ray diffraction measurements. Our DFT calculations and our model, based on the Landau theory of phase transitions, explain how the cooperative thermal switching of molecular spin state drives a ferroelastic symmetry breaking at high temperature, through a coupled Jahn-Teller distortion. Ferroelastic materials have rich properties, with important applications in memory, multifunctional and novel controllable devices. The electronic bistability in soft functional materials represents an important source of entropy gain, capable of overcoming the cost of symmetry-breaking entropy, which opens up new perspectives for stabilizing high-temperature and low-symmetry ferroic functions of advanced materials.

**New concepts**

In this manuscript, we present the concept of generating symmetry-breaking based functions of materials at high temperature, which is made possible by coupled electronic instability. Symmetry breaking is responsible for the emergence of functions associated with different types of ferroic orders, enabling applications in sensors, memories, transducers or actuators. However, symmetry breaking usually occurs during cooling, as phases with low symmetry are often those with low entropy. This fact severely limits the practical applications of many materials at room temperature. This study highlights the effectiveness of using an electronic bistability, corresponding here to spin-crossover, as a large source of entropy gain upon warming to overcome the entropy cost of a coupled symmetry breaking, corresponding here to a ferroelastic distortion. The importance of the coupling between both types of instabilities has been proven through a combination of magnetic, calorimetric and structural experiments and theoretical models. This communication thus demonstrates a proof-of-concept of ascending symmetry-breaking driven through an isosymmetric electronic instability, which opens the way to applications for functionalized materials operating at high temperature and more particularly at room temperature.

## 1. Introduction

The study of solid-state phase transitions lies at the very heart of materials science.[1] The associated changes in physical properties and functions, which can be driven by many external control parameters such as temperature, pressure, magnetic or electric field, and even light, give rise to various types of

[a.] Univ Rennes, CNRS, IPR (Institut de Physique de Rennes) - UMR 6251, 35000 Rennes, France.
[b.] CNRS, Univ Rennes, DYNACOM (Dynamical Control of Materials Laboratory) - IRL 2015, The University of Tokyo, 7-3-1 Hongo, Tokyo 113-0033, Japan.
[c.] Univ Rennes, Ecole Nationale Supérieure de Chimie de Rennes, CNRS, ISCR (Institut des Sciences Chimiques de Rennes) - UMR 6226, F-35000 Rennes, France
[d.] Univ. Bordeaux, CNRS, Bordeaux INP, ICMCB, UMR5026, F-33600 Pessac, France
[e.] Institut Universitaire de France (IUF).

† Supplementary Information available: Intra-molecular structural reorganization, Thermal dependence of lattice parameters and calculations of the volume strain and ferroelastic distortion, Symmetry-breaking Bragg peaks, X-Ray diffraction experimental details, DFT calculations, Landau model of coupled symmetry-breaking and SCO, Differential scanning calorimetry measurements. Supplementary video 1: Unstable torsion mode QT in the HShs state.
See DOI: 10.1039/x0xx00000x

applications for materials. In a large class of phase transitions, materials exhibit a symmetry-breaking,[2,3] where the associated ordering of structural and/or electronic degrees of freedom is responsible for the emergence of various functions related to ferromagnetism, ferroelectricity, ferroelasticity, multiferroicity, or conductivity.[4-12] According to thermodynamics, in the general situation of symmetry-breaking (SB), the high temperature (HT) phase always has higher entropy and higher symmetry than the low-symmetry and low temperature (LT) phase. These types of SB processes have been rationalized in the framework of the Landau theory of phase transitions,[2,3] where the LT phase results from the loss of some symmetry operators of the HT phase.

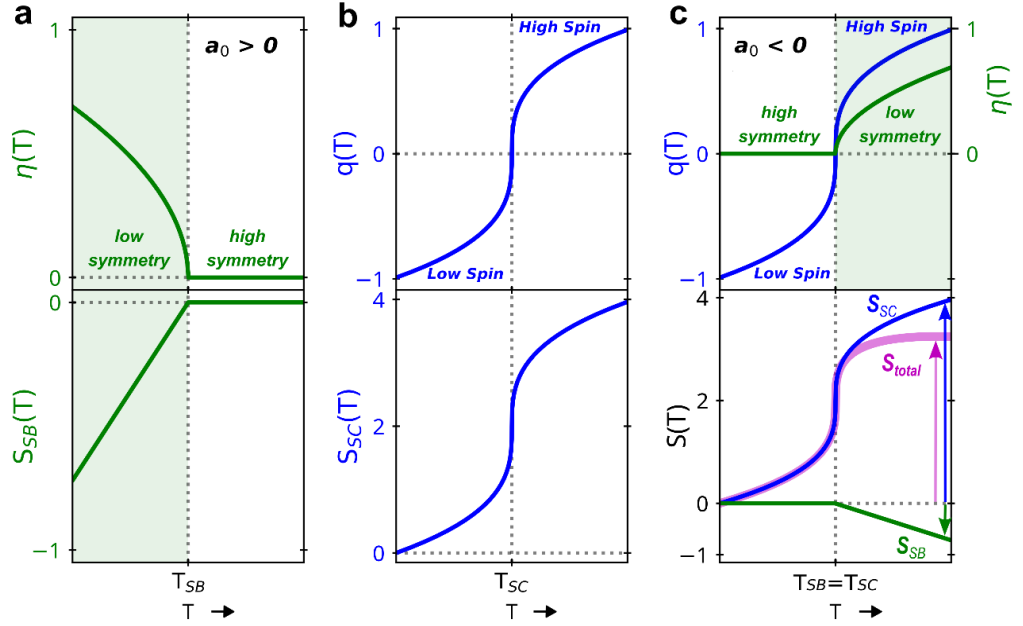

**Fig. 1** Basics of Landau theory. a) Usual descending symmetry-breaking (SB, $a_0 > 0$). The low-symmetry phase (green area) forms below $T_{SB}$, characterized by SB parameter $\eta(T)$ and SB entropy ($S_{SB}$). b) Spin crossover (SC) curve $q(T)$ driven by the entropy increase ($S_{SC}$) in high spin state. c) Concept of ascending symmetry-breaking ($a_0 < 0$) upon heating: the SB entropy cost $S_{SB}$ can be largely compensated by the SC entropy gain $S_{SC}$. The total entropy $S_{tot} = S_{SB} + S_{SC}$, increases then upon warming, stabilizing a high spin and low symmetry phase at high temperature.

The degree of SB below a critical temperature $T_{SB}$ is characterized by the order parameter $\eta$. In the simplest case, when $\eta$ is scalar and continuous symmetry breaking occurs, the SB Gibbs energy is expanded in a series of even powers of $\eta$:

$$G_{SB}(T,\eta) = \frac{a_0}{2}(T-T_{SB})\eta^2 + \frac{b}{4}\eta^4 \quad (1)$$

The stability conditions ($b > 0$, $\frac{dG_{SB}}{d\eta} = 0$, $\frac{d^2G_{SB}}{d\eta^2} > 0$) are balanced by the $\eta^2$ coefficient. For $a_0 > 0$, the high-symmetry phase ($\eta = 0$) is stable above $T_{SB}$. Below $T_{SB}$, the low-symmetry phase is characterized by two symmetry-equivalent solutions $\pm\eta = \sqrt{\frac{a_0}{b}(T_{SB}-T)}$, with $G_{SB}(T) = -\frac{1}{4b}[a_0(T_{SB}-T)]^2$ and the SB entropy $S_{SB} = -\frac{\partial G_{SB}}{\partial T} = \frac{a_0^2}{2b}(T-T_{SB})$. Fig. 1a illustrates this usual descending symmetry-breaking phenomenon, which stabilizes the low-symmetry phase below $T_{SB}$, as rationalized through the $\eta(T)$ and $S_{SB}(T)$ curves. In this way, the symmetry of a material and its entropy commonly increase upon warming, as $\eta$ decreases. Due to entropy gain as the temperature rises, the high-symmetry phase is the high entropy one.

Recently, a counter idea has been discussed[13, 14]: can the symmetry of a material decrease with increasing temperature? This situation corresponds to $a_0 < 0$ (Fig. 1c), which means that the SB entropy decreases above $T_{SB}$. Another source of entropy gain is then needed to comply with thermodynamics. Thus, few systems exhibit such unusual and counter-intuitive ascending symmetry-breakings. A famous example is $^3$He, which exhibits a liquid-to-solid phase transition upon warming.[15] It is due to the larger entropy in solid state, originating from disordered nuclear spins, compared to the quantum Fermi nature of the liquid state. Other ascending symmetry-breakings have attracted sustained interest in superconducting,[16] ferroelectrics[17] or magnetic[18] materials. However, the origin of the entropy raising, required to stabilize the low-symmetry phase at high temperature, is often difficult to identify.

Here, we report on an ascending ferroelastic symmetry-breaking in a molecular spin crossover material and provide a simple explanation of the entropy stabilization process.

Spin crossover (SC) materials[19-22] exhibit more or less cooperative thermal conversions of their molecular electronic state from Low-Spin (*LS*) to High-Spin (*HS*). This is monitored at macroscopic scale through the totally symmetric SC parameter $q = \frac{N_{HS}-N_{LS}}{N_{HS}+N_{LS}}$, where $N_{LS}$ and $N_{HS}$ refer to the number of molecules in each state. The spin crossover phenomenon is isosymmetric, *i.e.* it occurs without change of symmetry, like gas-liquid transitions. SC occurring around $T_{SC}$ is described in the simplest way[23, 24] through the spin crossover Gibbs energy:

$$G_{SC}(T,q) = a_1(T_{SC}-T)q + \frac{B}{2}q^2 + \frac{C}{4}q^4 \quad (2)$$

The first term accounts for the coupling of $q$ to the conjugated field. When elastic interactions are strong ($B<0$) *LS* and *HS* phases coexist around $T_{SC}$, while a crossover regime appears for $B>0$. For the intermediate critical regime ($B=0$), $T_{SC}$ corresponds to the critical temperature. Then, the stability conditions of Eq (2) ($a_1 > 0, C > 0, \frac{dG}{dq} = 0, \frac{d^2G}{dq^2} > 0$) give obvious analytical solutions:

$$q = [\frac{a_1}{c}(T-T_{SC})]^{\frac{1}{3}} \text{ and } G_{SC} = -\frac{3}{4}\left[\frac{[a_1(T_{SC}-T)]^4}{C}\right]^{\frac{1}{3}}.$$

The $q(T)$ plot in Fig. 1b shows the SC phenomenon from around $T_{SC}$, which is also an entropy-driven process.[19, 20] Indeed, the SC entropy increases from *LS* to *HS* states, with $S_{SC} = a_1\left[\frac{a_1(T-T_{SC})}{C}\right]^{\frac{1}{3}}$.

In several SC materials, spin crossover and symmetry-breaking phenomena compete.[23-25] In these SC materials, the coupling of the isosymmetric SC electronic bitability to the SB allows for controlling symmetry-related functions, such as conductivity, magnetic order, multiferroicity or giant magnetoelectric response, through spin states conversion upon thermal, optical or magnetic stimuli.[25-31] In particular, ferroelastic materials have rich properties, with important applications in memory, multifunctional and novel controllable devices.[4, 8-10] However, the ferroelastic phases usually form at low temperature, which limits further applications.

Here we present and discuss the unusual emergence of an ascending ferroelastic symmetry-breaking in the high temperature phase of the SC material [Fe$^{II}$(PM-PEA)$_2$(NCS)$_2$]. It exhibits a Low-Spin high-symmetry phase (*LShs*) at low temperature and a ferroelastic High-Spin low-symmetry phase (*HSls*), stable above 235 K. Our crystallographic and magnetic measurements highlight the coupling between SC and SB phenomena. DFT calculations reveal a Jahn-Teller instability in the High-Spin state, breaking molecular symmetry. Our Landau model, describing the effect of the coupling between the two clearly identified SC and SB instabilities, rationalizes the emergence of a low-symmetry high temperature phase. Fig. 1c schematically shows the concept, where the entropy cost due to symmetry-breaking at high temperature is largely compensated by the entropy gain due to spin crossover.

## 2. Materials and methods

### 2.1 Synthetic procedures

Our measurements were performed on black needle-shaped single crystals, obtained in methanol after a few days by slow diffusion of the iron solution through a methanol layer into the ligand solution, in a straight tube under an inert atmosphere. More details can be found in previous work.[32]

### 2.2 Magnetic measurements

Data were collected from one single crystal glued with cryogenic GE7031 varnish to a 3 mm wide 15 cm long 0.725 mm thick strip cut out of a 200 mm high purity CZ type Si wafer (Neyco) whose diamagnetic contribution has been previously recorded.

### 2.3 X-ray diffraction

X-ray diffraction measurements of the [Fe$^{II}$(PM-PEA)$_2$(NCS)$_2$] molecular material were performed as function of temperature on a single crystal from 80 to 300 K and powder from 300 to 520 K. Complete experimental details can be found in the supplementary sections 1-4 in ESI†.

### 2.4 DSC measurements

Differential scanning calorimetry measurements were performed in the 115–295 K temperature range under argon on a Perkin Elmer calorimeter DSC8000 coupled with CLN2 regulator setting the scan rate of 5 K.min$^{-1}$ on 8.034 mg of crystals (supplementary section 7 in ESI†).

### 2.5 Computational Details

DFT geometry optimizations were carried out by using the Gaussian 16 (revision A.03) package[33] with the PBE0 hybrid functional[34, 35] and tightening both self-consistent field (10$^{-10}$ au) and geometry optimization (10$^{-5}$ au) convergence thresholds (supplementary section 5 in ESI†).

## 3. Results and discussion

### 3.1 Spin crossover

The [Fe$^{II}$(PM-PEA)$_2$(NCS)$_2$] compound exhibits a spin-transition with a thermal hysteresis loop.[32] Fig. 2a shows the temperature dependence of its $\chi_M T$ product, where $\chi_M$ is the molar magnetic susceptibility and T is the temperature. $\chi_M T$ directly relates to the SC parameter $q$ (right axis), through the weighted contribution of the number $N_{HS}$ of paramagnetic HS sites ($\chi_M^{HS}$ with S=2) and $N_{LS}$ of diamagnetic LS sites ($\chi_M^{LS}$ with S=0). In the HT phase, $\chi_M^{HS} T \simeq 3.5$ cm$^3$.K.mol$^{-1}$, which corresponds to a fully HS state ($q = 1$). As temperature is lowered, $q$ drops around $T_\downarrow \simeq 219\ K$ in the LT diamagnetic phase and gradually reaches a fully LS macroscopic state below 150 K. The characteristic $\chi_M T$ value close to 0 corresponds to $q = -1$. In contrast to polycrystalline powder samples,[32, 36] single-crystals exhibit a 16 K wide thermal hysteresis, with a clear single-step conversion upon cooling and two different conversion regimes upon heating: the gradual one from 150 K to $T_\uparrow \simeq 235\ K$, followed by a discontinuous jump towards the HS state. It was shown recently that contrary to isosymmetric SC, with symmetric conversion curves around $T_{SC}$ (Fig. 1b), such unsymmetric hysteresis loops are characteristic of the coupling between symmetry-breaking and spin crossover.[23]

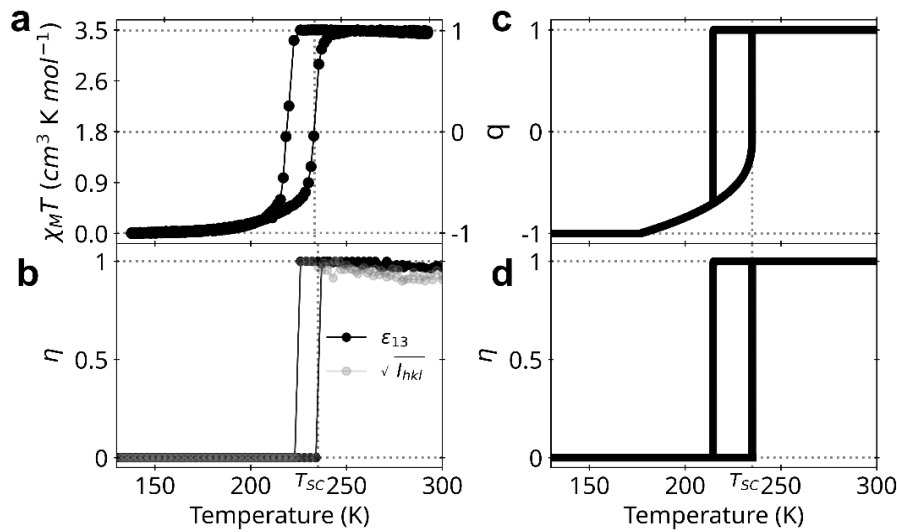

**Fig. 2** Coupled spin transition and ferroelastic distortion. a) Thermal dependence of the spin transition order parameter $q(T)$, monitored through $\chi_M T = \left[\left(\frac{1+q}{2}\right)\chi_M^{HS} + \left(\frac{1-q}{2}\right)\chi_M^{LS}\right] \times T$. b) Thermal dependence of the symmetry-breaking parameter $\eta(T)$ measured through the ferroelastic distortion $|\varepsilon_{13}|$ and the intensity of symmetry-breaking Bragg peaks ($\eta \propto \sqrt{I_{hkl}}$), both normalized to unity for clarity. All measurements are performed on a single crystal. c-d) Corresponding thermal dependence of SB and SC parameters obtained from our Landau model based on Eq(4).

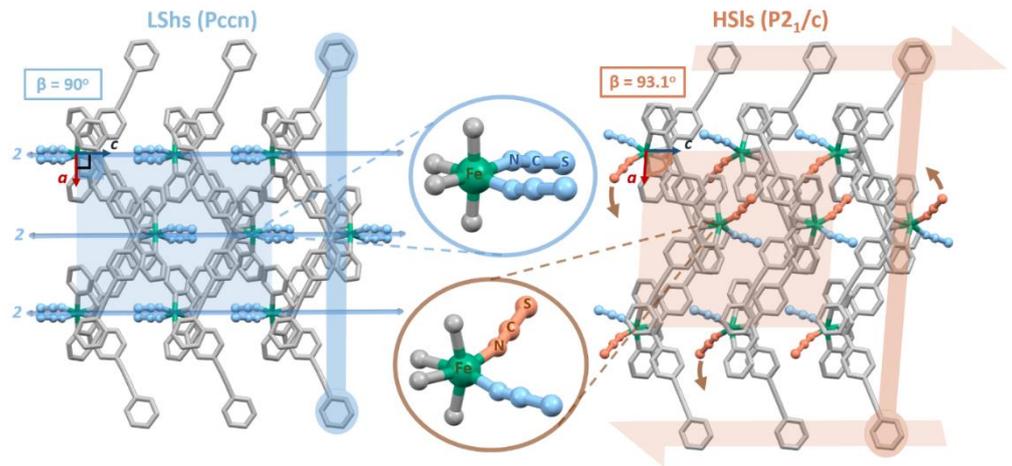

**Fig. 3** Symmetry-breaking ferroelastic transition. Change of crystalline symmetry from the LT parent-paraelastic *LShs* Pccn phase to the HT daughter-ferroelastic *HSls* $P2_1/c$ phase. The 2-fold molecular axes (2) along *c* in the *LShs* state are lost in the *HSls* phase, which is associated with a shear of the molecular layers (orange arrows) opening the β angle, and an asymmetric torsion of the PM-PEA ligands and thiocyanate groups (NCS zoom).

### 3.2 Ferroelastic symmetry-breaking at high temperature

From previous reports on [Fe$^{II}$(PM-PEA)$_2$(NCS)$_2$],[32] it was unclear whether the change between orthorhombic *LS* and monoclinic *HS* space groups corresponded to a reconstructive process[37] or to a group-subgroup symmetry-breaking.[3] However, this is a key point for its analysis in the framework of Landau theory, which only accounts for the latter case.[2] To learn more about the structural reorganization associated with the spin transition, we collected X-ray diffraction data on a single crystal upon cooling and heating. The changes in lattice parameters and Bragg peaks intensities (Supplementary Section 3, ESI†) reveal that the change of spin state is concomitant with a space group change from the *Pccn LS* LT phase to the $P2_1/c$ *HS* HT phase. This ferroelastic phase transition, characterized by the deviation of the lattice $\beta$ angle from 90° to 93.1°, corresponds to a symmetry-breaking with a group-sub-group relationship between the *Pccn* space group of the parent-paraelastic *LShs* and the $P2_1/c$ space group of the daughter-ferroelastic *HSls* phase (Fig. 3).[38-40] This long-range ferroelastic distortion is monitored through the amplitude of the symmetry-breaking spontaneous strain: $|\varepsilon_{13}(T)| = |\frac{1}{2}\left(\frac{c(T)\cos(\beta(T))}{c_0(T)}\right)|$.[4, 38, 40] In principle, a continuous *Pccn*↔$P2_1/c$ phase transition line can exist. However, Fig. 2 shows that $|\varepsilon_{13}(T)| \propto \eta(T)$, referred to as the symmetry-breaking order parameter hereafter, changes discontinuously at $T_\downarrow \simeq 221\ K$ and $T_\uparrow \simeq 235\ K$ (Fig. 2b). $\eta(T) = 0$ in *LShs* phase and $\eta(T)$ is almost constant in the *HSls* phase from 221 to 300 K. The analysis of powder diffraction patterns indicates that the daughter and high temperature ferroelastic *HSls* phase, characterized by the splitting of $(hkl)$ and $(\bar{h}kl)$ Bragg peaks, remains stable up to $\simeq$500 K, beyond which the sample melts (Fig. S5, ESI†). This ascending SB from LT to HT phases is also characterized by the change of Bragg peaks extinction rules (Supplementary Section 4, ESI†). The intensities $I_{hkl}$ of the $(0kl)$ Bragg peaks with $l = 2n + 1$ and $(hk0)$ Bragg peaks with $h + k = 2n + 1$ are zero in the *Pccn LShs* phase and non-zero in the $P2_1/c$ *HSls* phase. Fig. 2b shows the thermal dependence of $\sqrt{I_{hkl}(T)} \propto \eta(T)$ for these symmetry-breaking Bragg peaks, confirming the discontinuous symmetry change between the LT *LShs* to the HT *HSls* phases.

The lattice parameters (Fig. S3, ESI†) also exhibit discontinuous changes upon heating and cooling during the phase transition, resulting in a large volume strain $v_s \simeq 0.05$ in the *HSls* phase. The thermal evolutions of the spin transition parameter $q$, lattice parameters, volume strain $v_s$, intensity of symmetry-breaking Bragg peaks, and ferroelastic distortion $\varepsilon_{13}$, exhibit the same thermal hysteresis between *LShs* and *HSls* phases (Fig. 2). It clearly characterizes the coupling between SC and SB phenomena, driven by long-range elastic interactions.[23]

Fig. 3 shows the crystalline structure of [Fe$^{II}$(PM-PEA)$_2$(NCS)$_2$] along the *b* axis of the lattice in the fully *LS* (90 K) and *HS* (300 K) states. The crystal packing exhibits a ferroelastic distortion, corresponding to a shear of the molecular layers in the *HSls* phase. The resulting deviation of the lattice angle β from 90° to 93.1° is related to the loss of the C2 symmetry. In the *Pccn LShs* space group, the molecules are located on Wyckoff positions with C2 symmetry along *c* molecular axis, while in the $P2_1/c$ *HSls* phase, the C2 symmetry is lost, with distorted molecules in general positions. This is characterized by an important torsion of the thiocyanate groups and a moderate rearrangement of the PM-PEA ligands in the *HSls* state, initially equivalent in the *LShs* state. The intra-molecular structural parameters,[41] such as the <Fe-N> bond lengths and N-Fe-N torsion angles, are characteristic of *LS* and *HS* states (Supplementary Section 2, ESI†). Fig. S6 shows that the monoclinic *HSls* state can also be quenched by flash cooling at 30 K, while the evolution of $v_s$ and $\varepsilon_{13}$ indicate that the relaxation towards the orthorhombic *LShs* ground phase occurs around 65 K upon heating. These results underline the strong coupling between the molecular spin state ($q$) and the crystalline and molecular symmetries ($\eta$).

Then, comes the chicken-egg question: is the spin crossover responsible for symmetry-breaking, or vice versa?

### 3.3 DFT analysis of the Jahn-Teller distortion in the high spin state

For understanding the coupled molecular changes of spin state and symmetry, we performed DFT calculations on the isolated molecule (Supplementary Section 5, ESI†). In the *LS* state, we optimized the molecular structure with C2 point symmetry of the Wyckoff position of the molecular sites in the *LShs* lattice.

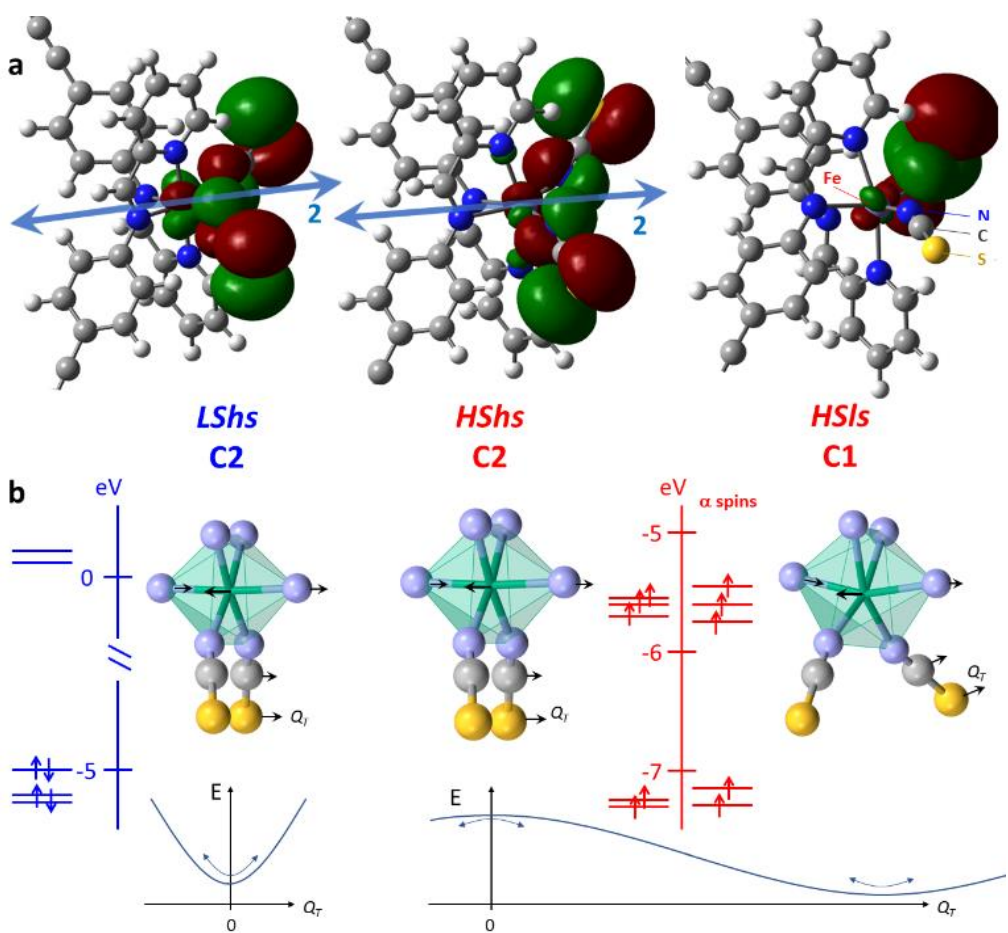

**Fig. 4** Comprehensive summary of DFT calculations. a) Optimized molecular structure for C2 *LShs*, C2 *HShs* and C1 *HSls* states. The loss of the 2-fold molecular axis in the *HSls* state with the molecular torsion results in asymmetric orbitals, especially around the NCS groups. b) Zoom on the $FeN_4(NCS)_2$ core, with energy diagram of $t_{2g}$-like and $e_g$-like orbitals showing the splitting of orbital degeneracy from C2 (*LShs* and *HShs*) to C1 (*HSls*), where only α spins are shown for clarity. The C2 *HShs* state is unstable along one torsion mode $Q_T$ (B symmetry), which has a negative frequency. The main atomic motions for $Q_T$ (see supplementary video 1) are schematically represented by arrows for the atoms surrounding the Fe. This Jahn-Teller distortion breaks C2 symmetry and brings the molecule towards the stable *HSls* state with C1 symmetry.

The results provide positive frequencies for all the 231 intramolecular vibrational modes with A or B symmetry. This indicates that the molecular structure in the *LShs* state is stable with respect to the C2 symmetry, with symmetry-equivalent thiocyanate groups and PM-PEA ligands (Fig. 4). The energy diagram of the molecular orbitals in the *LShs* and *HShs* states with C2 symmetry shows the degeneracy of some orbitals with symmetric thiocyanate groups. In contrast, the structure of the *HShs* state with C2 symmetry is unstable, as characterized by the negative frequency of one torsion mode $Q_T$ (B symmetry, Fig. S9 and Supplementary video 1, ESI†).

The molecular torsion along $Q_T$ stabilises the HS state, as schematically shown in Fig. 4, which lowers molecular symmetry towards C1. The optimized molecular structure in the *HSls* state with C1 symmetry is stable (all modes have positive frequencies), and the structural distortion of the thiocyanate groups correlates with their asymmetric spin density in the *HSls* state. The overall molecular distortion in the *HSls* state provided by DFT calculations is in good agreement with crystallographic data. These calculations reveal the microscopic origin of the symmetry-breaking, due to the Jahn-Teller distortion along $Q_T$ in the *HSls* state, characterized by an asymmetric bending of the $Fe^{2+}-(NCS^-)_2$ dipoles as the 2-fold molecular axis is lost.

Our DFT results underline again the strong coupling between spin state and symmetry, as the Jahn-Teller distortion stabilizes the High-Spin state.

### 3.4 Coupled symmetry-breaking and spin crossover

Describing the *LShs* to *HSls* phase transition requires considering the contributions of both the SB (Eq. 1) and SC (Eq. 2) phenomena to the Landau thermodynamical potential as well as their coupling terms. This approach was used for other materials[23, 42] to describe their phase diagrams and explain why SC and SB phenomena can occur sequentially or simultaneously. Hereafter we use the symmetry-relevant potential:

$$G(q,\eta) = a_1(T_{SC} - T)q + \frac{B}{2}q^2 + \frac{C}{4}q^4 + \frac{a_0}{2}(T - T_c)\eta^2 + \frac{b}{4}\eta^4 + Dq\eta^2 + \frac{E}{2}q^2\eta^2 \quad (3)$$

The couplings ($D < 0$ and $E < 0$) contribute to stabilizing the *LShs* ($q = -1$, $\eta = 0$) and the *HSls* ($q = 1$, $\eta \neq 0$) phases revealed by experiments and DFT calculations. It should be underlined that, compared to other studies related to "improper" linear-quadratic coupling considered in the literature,[37, 43] here it is the inclusion of the symmetry-allowed $q$ term that balances the relative stability of HS and LS states through their entropy difference. It was shown that this phenomenological coupling includes both the elastic energy and elastic coupling of both parameters to lattice strains, which is responsible for cooperative elastically-driven phase transitions.[23, 42] In order to further simplify the model, we consider the intermediate critical crossover case (B=0), and we use the strong coupling limit (Supplementary Section 6, ESI†), for which $\frac{a_0}{2}(T - T_c) \ll Dq$. Then the potential reduces to:

$$G(q,\eta) = a_1(T_{SC} - T)q + \frac{C}{4}q^4 + \frac{b}{4}\eta^4 + Dq\eta^2 + \frac{E}{2}q^2\eta^2 \quad (4)$$

The equilibrium $(q,\eta)$ values are found by minimizing the Gibbs energy in equation (4). $\eta = 0$ is stable for $q < 0$, which corresponds to the *LShs* phase. For $q > 0$, the two symmetry-equivalent stable solutions, $\pm\eta = \sqrt{-(2Dq + Eq^2)/b}$, correspond to the two possible ferroelastic domains in the *HSls* phase upon the $Pccn \rightarrow P2_1/c$ symmetry breaking.[38-40] Fig. 5 shows how the equilibrium $(q,\eta)$ values change with temperature, considering the spin transition parameter as explicitly restricted to $-1 \leq q \leq 1$. Figs. 2c-d show the $q(T)$ and $\eta(T)$ curves calculated from equation (4). These are in very good agreement with the experimental data, as they reproduce the main experimental features of the phase transition. i) On heating from the fully *LShs* state ($q = -1, \eta = 0$) a SC starts as $q$ gradually increases (Fig. 2a & 2c). ii) Above the critical temperature $T_\uparrow$, the *LShs* phase is no more stable and a discontinuous transition occurs towards the *HSls* phase, where $q > 0$ is stabilized by the SB ($\eta \neq 0$). iii) The couplings maximize $q$ to 1, which corresponds to the fully HS state. The amplitude of the SB parameter is then constant ($\eta = \sqrt{-(2D + E)/b}$) in the *HSls* phase (Fig. 2b & 2d). vi) The coupling terms generate an energy barrier in the potential, which stabilizes the *HSls* ($\Delta G_{HL}$, Fig. 5) and precludes gradual conversion towards LS on cooling. The *HSls* state is then stable down to $T_\downarrow$, where the *LShs* phase is reached. The potential in equation (4) differs therefore in the low-symmetry ($\eta \neq 0$) and high-symmetry ($\eta = 0$) phases and the coupling opens a thermal hysteresis for both $q$ and $\eta$.

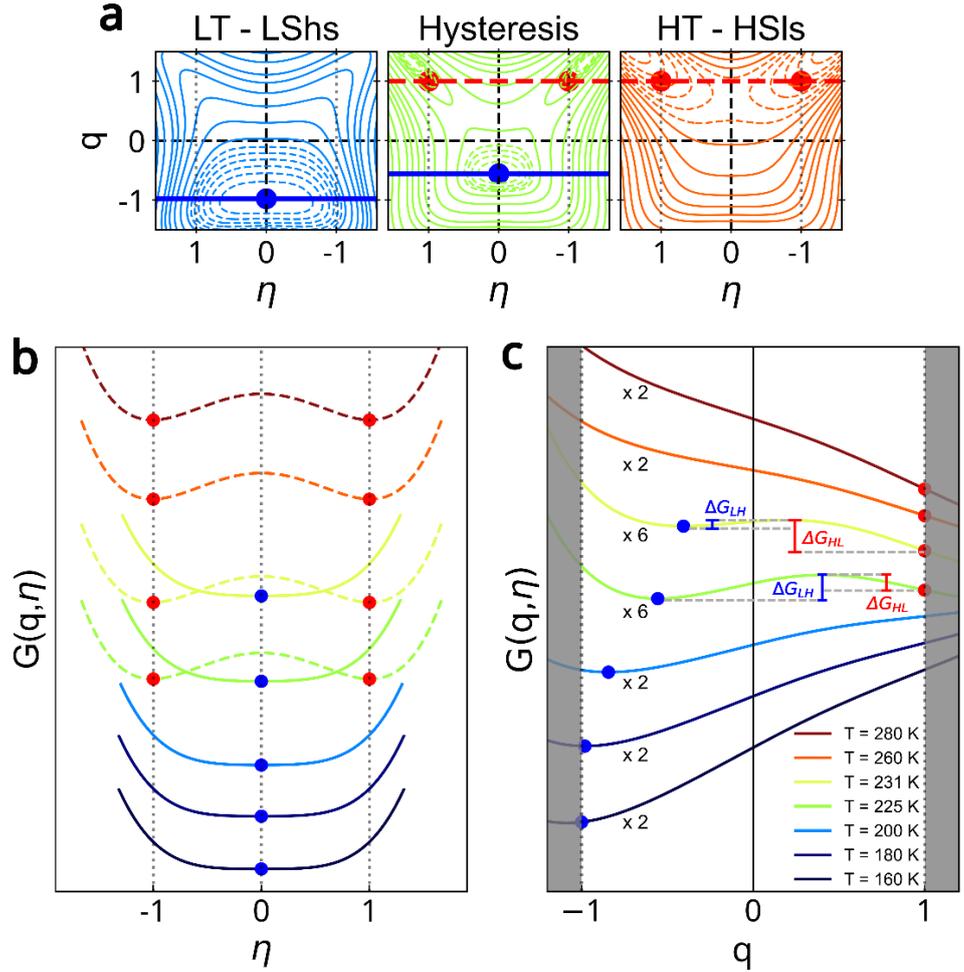

**Fig. 5** Coupled spin crossover and symmetry-breaking. a) Thermal dependence of the Gibbs energy $G(q,\eta)$ in LT *LShs* phase ($q = -1, \eta = 0$ left), HT *HSls* phase ($q = -1, \eta \neq 0$ right) and inside hysteresis (middle). b) Cut of $G(q,\eta)$ along the SB order parameter ($\eta$) and SC order parameter ($q$) axes. $q$ saturates to $-1$ at LT and gradually increases on approaching $T_\uparrow$, where it discontinuously jumps and saturate to 1 at HT. In the *LShs* phase $\eta = 0$, while in the *HSls* phase $\eta = \pm\sqrt{-(2D + E)/b}$, with $|\eta|$ normalized to 1 for clarity. The coupling stabilizes $q = 1$ on cooling down to $T_\downarrow$ as an energy barrier appears from LS to HS ($\Delta G_{LH}$) and HS to LS ($\Delta G_{HL}$) inside thermal hysteresis (curves are rescaled by 2 or 6 for a better visualization). The results are obtained from Eq (4) with $a_1 = 0.6, C = 35, b = 10, T_{SC} = 235\ K, D = -6.5, E = -12.2$.

This cooperative and hysteretic phase transition is mediated by the elastic interactions within the molecular lattice. The emergence of a ferroelastic ordering in the high temperature phase of a SC material, from 235 K to 500 K beyond which the material melts, offers new physical perspectives for ascending symmetry-breaking. Heat capacity measurements reveal large entropy changes between the *HSls* and *LShs* phases during the phase transition upon heating ($\Delta S = 62.8\ J.mol^{-1}.K^{-1}$) and and cooling ($\Delta S = 67.6\ J.mol^{-1}.K^{-1}$, Supp. Section 7 in ESI†). These values are within the usual 40–80 $J.mol^{-1}.K^{-1}$ range for SC materials.[19] However, the entropy increase from *LShs* to *HSls* phases includes contributions from both SB and SC phenomena: $\Delta S_{tot} = \Delta S_{SB} + \Delta S_{SC}$.

On the one hand, the symmetry-breaking entropy $\Delta S_{SB}$ is negative due to ascending symmetry-breaking. Indeed, if the ascending orthorhombic → monoclinic SB was purely order-disorder in character,[44] with molecules ordered on a single position in HT phase and disordered between two positions equivalent by C2 axis in LT phase, the SB entropy would be $\Delta S_{SB} = R\ln(1) - R\ln(2) = -5.76\ J.K^{-1}mol^{-1}$.

On the other hand, the SC phenomenon is a large source of entropy gain, responsible for giant barocaloric effect,[19, 45] with two main contributions: $\Delta S_{SC} = \Delta S_{spin} + \Delta S_{vib}$. $\Delta S_{spin}$ is due to different spin multiplicities ($\Omega = 2S + 1$) in HS ($S = 2$) and LS ($S = 0$) states: $\Delta S_{spin} = R(ln(5) - ln(1))$ = 13.42 J.K$^{-1}$.mol$^{-1}$. The vibrational entropy difference $\Delta S_{vib}$ is mainly due to the lower frequencies of the FeN$_6$ core vibrations in the less bonding HS state compared to LS.[19, 20] In the present case, we estimate for [Fe$^{II}$(PM-PEA)$_2$(NCS)$_2$] $\Delta S_{vib}$ = 58.9 J.K$^{-1}$.mol$^{-1}$, based on our DFT calculations of the vibrational modes in both spin states (Supplementary Section 5, ESI†). In this way, we calculate a global entropy increase upon heating from *LShs* to *HSls* phases, which is mainly due to the change of electronic (spin) state: $\Delta S_{tot} = \Delta S_{SB} + \Delta S_{spin} + \Delta S_{vib}$ = 66.5 J.K$^{-1}$mol$^{-1}$. This estimation is in nice agreement with our calorimetric measurements ΔS =62.8 – 67.6 J.mol$^{-1}$.K$^{-1}$ (Fig. S12, ESI†). Thus, the molecular scale plays a central role in such materials, both for entropy stabilization, through numerous intra-molecular degrees of freedom, and for the Jahn-Teller distortion associated with the ferroelastic symmetry breaking.

Fig. 1c represents schematically this concept of ascending symmetry-breaking, where the SB entropy cost ($\Delta S_{SB}$) can be largely compensated by the isosymmetric entropy gain ($\Delta S_{SC}$) due to electronic bistability. However, a rigorous description of the concerted SC and SB phenomena requires considering the important role of their coupling, as explicitly discussed through our Landau model (Fig. 2 and Fig. 5). This coupling, which strongly stabilizes the *HSls* state, also plays a key role for the stabilization of *HSls* states trapped by flash cooling (Fig. S6, ESI†) or photoinduced.[46] Overall, the *LShs* → *HSls* ascending symmetry-breaking reported here corresponds to an entropy-driven process, where the thermal SC starting upon warming drives the cooperative ferroelastic distortion above a critical conversion. This ascending symmetry breaking is the consequence of the collective distortions of the molecules in the lattice, due to cooperative elastic interactions.

We can notice that a hypothetical parent para-elastic phase of high spin and high-symmetry ($Pccn$) could form around 800 K, as suggested by experimental data (see Fig. S5, ESI†), and our Landau model when the strong coupling limit vanishes (Fig. S11, ESI†). However, this *HShs* phase is not observed as the sample melts around 500 K.

## 4. Conclusions

Ascending symmetry-breaking is often associated with re-entrant phase transitions, where the low-symmetry phase is intermediate between two high-symmetry LT and HT phases. In such a case, symmetry-breaking takes place over a narrow temperature range and results from competing interactions or frustrations.[16, 18, 47, 48] This is also the case for some stepwise spin transitions with long-range *HS-LS* order. In this case, the intermediate half-conversion phase, exhibiting symmetry-breaking ($q = 0$, $\eta \neq 0$) is stabilized by a bi-quadratic $q^2\eta^2$ coupling over few Kelvins, in between high-symmetry *LS* and *HS* phases.[24, 25, 30, 42, 49] Instead, our study rationalizes how a symmetry-broken phase can be stabilized, through an isosymmetric electronic bistability, over a broader and higher temperature range, here from 235 K and well above room temperature. Compared with hard condensed matter, soft SC molecular materials exhibiting electronic bistability hold a large source of vibrational entropy gain. This substantial advantage opens up fascinating prospects for stabilizing robust low-symmetry phases in materials, enabling so the development of future generations of room-temperature ferroic devices, such as switches, force sensors and non-volatile memories.[4, 50, 51]

Regarding the ferroelastic domains forming after thermal cycling, it is well-established now that domain walls form 2D functional objects in their own right, which can be engineered to give rise to an electric polarization in an otherwise centrosymmetric system or to pyroelectric and electrocaloric effects in ferroelectric.[10, 30, 52] Regarding bulk properties, it was shown that an electric field can be used to substitute non-polar ferroelastic phases[53] exhibiting an antiferroelectric array of dipoles (such as the Fe$^{2+}$–(NCS$^-$)$_2$ bonds in Fig. 3), by a polar phase. The interesting thing about the single domain nature of the ferroelastic crystals [Fe$^{II}$(PM-PEA)$_2$(NCS)$_2$], synthesized at room temperature, is that they can give rise to a single polar domain under the effect of an electric field.

The description of symmetry-breaking in the framework of the Landau theory of phase transitions is universal. Our model, considering a coupled electronic bistability successfully reproduced phase diagrams and the occurrence of simultaneous or sequential electronic bistability and SB phenomena in a broad variety of system.[23, 42] The present concept of ascending symmetry-breaking coupled to an electronic instability can also apply to ascending ferroelectricity or ferromagnetism, as long as the associated electric or magnetic orders couple to an isosymmetric electronic bistability, source of large entropy gain. This concept opens the way to applications for materials functionalized by their low symmetry and operating at high temperature.

## Author contributions

F.J.V.M.: methodology, data curation, formal analysis, writing – original draft; R.G.T.R.: conceptualization, data curation; E.T.: supervision, methodology, data curation, formal analysis; T.B.: data curation; H.C.: conceptualization, & review; N.D.: sample synthesis; D.D: data curation; P.G.: data curation; G.C.: data curation; B.L.G.: DFT calculations; E.C.: Funding acquisition, data curation, investigation, conceptualization, supervision, writing – original draft. All authors critically reviewed and revised the manuscript draft and approved the final version for submission.

## Conflicts of interest

There are no conflicts to declare.

## Data availability

Crystallographic data has been deposited at the CCDC. The CCDC 2309697 (90 K) and CCDC 2309698 (300 K) contain the crystal data collection and refinement parameter details for this paper, which can be obtained free of charge via www.ccdc.cam. ac.uk/conts/retrieving.html (or from the Cambridge Crystallographic Data Centre, 12 Union Road, Cambridge CB2 1EZ, U.K.; fax: (+44) 1223-336-033; or deposit@ccdc.ca.ac.uk).

## Acknowledgements

This work was supported in part by Agence Nationale de la Recherche for financial support under the grant ANR-19-CE30-0004 ELECTROPHONE, ANR-19-CE07-0027 SMAC. EC thanks the University Rennes and the Fondation Rennes 1 for funding. F. J. V.-M. acknowledges the support of the European Social Fund (ESF) and Generalitat Valenciana for his postdoctoral fellowship (APOSTD/2021/359) and Région Bretagne (BIENVENÜE-MSCA COFUND n° 899546). The authors thank Région Bretagne, Ille-Et-Vilaine Department, French Ministry of Research, Rennes Métropole, CNRS, European Union for financial support (CPER Project Mat&Trans 2021-2027). The French GENCI/IDRIS-CINES centers for high-performance computing resources are also acknowledged.